\documentstyle[aas2pp4]{article}	

\lefthead{Sakamoto et al.}
\righthead{Bar-driven gas transport in galaxies}

\newcommand{\etal}{\mbox{\it et al.}}
\newcommand{\Lsol}{\mbox{$L_\odot$}}
\newcommand{\Msol}{\mbox{$M_\odot$}}
\newcommand{\Msolyr}{\mbox{$M_\odot$ yr$^{-1}$}}
\newcommand{\kms}{\mbox{km s$^{-1}$}}
\newcommand{\beam}{\mbox{beam$^{-1}$}}

\newcommand{\Xco}{$X_{\rm CO}$}
\newcommand{\fcon}{$f_{\rm con}$}
\newcommand{\Lfir}{$L_{\rm FIR}$}
\newcommand{\Mdyn}{\mbox{$M_{\rm dyn}$}}
\newcommand{\Mgas}{\mbox{$M_{\rm gas}$}}
\newcommand{\HH}{\mbox{H$_2$}}

\newcommand{\Sgascent}{\mbox{$\Sigma_{\rm gas}(R<500{\rm pc})$}}
\newcommand{\Sgasdisk}{\mbox{$\Sigma_{\rm gas}(R< R_{25})$}}
\newcommand{\Halpha}{\mbox{H$\alpha$}}
\newcommand{\Hbeta}{\mbox{H$\beta$}}
\newcommand{\nd}{\nodata}
\newcommand{\tm}[1]{\tablenotemark{#1}}

\begin{document}

\title{Bar-driven Transport of Molecular Gas to Galactic Centers \\
	and Its Consequences}

\author{
K. Sakamoto\altaffilmark{1,2}, 
S. K. Okumura\altaffilmark{1}, 
S. Ishizuki\altaffilmark{3} \\
\and
N. Z. Scoville\altaffilmark{2}
}

\altaffiltext{1}{Nobeyama Radio Observatory, Minamisaku, Nagano, 
384-1305, Japan; E-mail(KS): sakamoto@nro.nao.ac.jp}
\altaffiltext{2}{California Institute of Technology, Pasadena, CA 91125}
\altaffiltext{3}{National Astronomical Observatory, Mitaka, Tokyo, 
181-8588, Japan}

\vspace{-1cm}

\begin{abstract}
{\normalsize
	We study the characteristics of molecular gas in the central
regions of spiral galaxies on the basis of 
our CO(J=1--0) imaging survey of 20 nearby spiral galaxies 
using the NRO and OVRO millimeter arrays. 
Condensations of molecular gas at galactic
centers with sizescales $\lesssim 1$ kpc and CO-derived masses 
$M_{\rm gas}(R<500{\rm pc}) \sim 10^{8}$ -- $10^{9}$ \Msol\ are
found to be prevalent in the gas-rich $\sim L^{*}$ galaxies.
Moreover, the degree of gas concentration to the central kpc
is found to be higher in barred systems than in unbarred galaxies.
This is the first statistical evidence for the higher central
concentration of molecular gas in barred galaxies, and it strongly supports 
the theory of bar-driven gas transport.
It is most likely that more than half of molecular gas within the central kpc
of a barred galaxy was transported there from outside by the bar.
The supply of gas has exceeded the consumption of gas by star formation
in the central kpc, resulting in the excess gas in the centers of
barred systems.
The mean rate of gas inflow is statistically estimated to be larger than
0.1 -- 1 \Msolyr.
	
	There is no clear correlation between gas mass in the central 
kpc and the type of nuclear spectrum (HII, LINER, or Seyfert), 
suggesting that the amount of gas at this scale does not determine the 
nature of the nuclear activity. 
There is, however, a clear correlation for galaxies with larger 
gas-to-dynamical mass ratios to have HII nuclear spectra, 
while galaxies with smaller ratios show spectra indicating AGN. 
This trend may well be related to the gravitational stability 
of the nuclear gas disk, 
which is generally lower for larger gas mass fractions. 
It is therefore possible that all galaxies have active nuclei, 
but that dwarf AGN are overwhelmed by the surrounding star formation 
when the nuclear molecular gas disk is massive and unstable.

	The theoretical prediction of bar-dissolution by condensation
of gas to galactic centers is observationally tested by comparing
gas concentration in barred and unbarred galaxies.
If a bar is to be destroyed so abruptly that the gas condensation at the
nucleus does not have enough time to be consumed, then there
would be currently unbarred but previously barred galaxies with
high gas concentrations.
The lack of such galaxies in our sample, together with the current  
rates of gas consumption at the galactic centers, 
suggests that the timescale for bar dissolution is 
larger than $10^{8}$ -- $10^{10}$ yr or a bar in a $L^{*}$ galaxy
is not destroyed by a condensation of $10^{8}$ -- $10^{9}$ \Msol\ 
gas in the central kpc. 
}
\end{abstract}


\vspace{5mm}
\begin{center}
{\large
 To appear in {\it ApJ.}, vol. 525, No. 2, (10 Nov. 1999) issue.
}
\end{center}

\section{Introduction \label{s.intro}}
		Stellar bars in disk galaxies have long been recognized as 
powerful tools to transport interstellar gas toward galactic centers
(e.g., \cite{Matsuda77}; Simkin, Su, \& Schwarz 1980).
Gravitational torques from bars are much more efficient in gas transport
on galactic scale than viscous torques (Combes, Dupraz, \& Gerin 1990).
Typically gas inflow rates of the order of 0.1 -- 10 \Msolyr\ have been 
obtained in numerical simulations of barred galaxies with
ordinary amount of gas (e.g., \cite{FriedliBenz93}).

	Two consequences of the bar-driven 
gas transport have been extensively investigated; fueling to 
(circum)nuclear star formation and to active galactic nuclei (AGN), 
and secular evolution of galaxies involving bar dissolution and bulge 
growth.
Bursts of star formation in galactic centers 
($\lesssim 1$ kpc) consume $10^{7}$ -- $10^{9}$ \Msol\ of gas in
$10^{7}$ -- $10^{8}$ years (\cite{Balzano83}), and thus need
the supply of gas as bars can provide.
Active nuclei in barred galaxies may also benefit from 
the larger supply of infalling gas, though additional mechanisms may 
be necessary to deliver the gas from $R \lesssim 1$ kpc to the nucleus 
(Shlosman, Frank, \& Begelman 1989).
Another aspect of the bar-driven gas transport is a 
significant mass transfer in galactic disks,
possibly followed by drastic changes in the dynamical structure of galaxies.
It is predicted that bars are 
destroyed if the bar-driven mass concentration at the nuclei 
is large enough, and that some of disk stars are added to bulges as a 
result of the bar dissolution (\cite{Hasan90}; \cite{Pfenniger90}).
It has been inferred therefore that the formation-dissolution 
cycle of  bars forces spiral galaxies to evolve from late to early 
Hubble types, in the direction of increasing bulge-to-disk ratios
(\cite{FriedliMartinet93}; Hasan, Pfenniger, \& Norman 1993).

		On the observational side, there are three major pieces 
of evidence for bar-driven gas transport;  metallicity gradients in 
galactic disks, estimates of gas inflow rates in a few barred 
galaxies, and statistics on \Halpha\ luminosity (and other tracers of 
star formation) in galactic centers.
The radial gradients of metallicity in galactic disks
are shallower in barred galaxies than in unbarred galaxies 
(\cite{Costas92}; Zaritsky, Kennicutt, \& Huchra 1994; \cite{Martin94}).
The shallower gradients in barred systems 
are attributed to better mixing of interstellar medium by bars.
Metallicity gradients tell time-averaged effects of the 
bar-driven gas transfer, because the gradients are over galactic 
disks and hence vary on timescales larger than the dynamical 
time of the disks.

	Estimates of gas inflow rate in barred galaxies 
have been obtained using near-IR and/or CO observations, 
but only in a few objects. 
Quillen \etal\ (1995) constructed a mass model of the bar in NGC 7479
from optical and near-IR images, calculated the torques to be exerted 
on the molecular gas seen in CO, and estimated a mass inflow rate.
Benedict, Smith, \& Kenney (1996) detected in CO large inward velocities
at two positions near the center of a barred galaxy NGC 4314 
and suggested a gas inflow, though the proof of net inflow
was not provided.
Regan, Vogel, \& Teuben (1997) compared the \Halpha\ velocities 
in a barred galaxy NGC 1530 with hydrodynamical models
and gas distributions to calculate the mass inflow rate.
The estimated rates of gas flow are of the order of 0.1--10 \Msolyr\ 
all directed inward.
There is, however, a fundamental difficulty in the observational 
estimation of mass inflow.
The gas flux into a region is a line integral 
(on the boundary of the region) of gas mass multiplied by 
inward velocity normal to the boundary.
The inward velocities are not observable on 
the minor axis of a galaxy, and thus
the inflow rate can not be usually measured without 
a gas dynamics model (e.g., Regan \etal\ 1997).
The model usually needs multi-wavelength observations
and assumptions on the mass-to-luminosity ratio, 
gas properties, the pattern speed of the bar, etc.
Statistics on the gas inflow rates have not been 
obtained in this way, presumably owing to the complexity of the method. 
The estimated rates of gas inflow are current and 
instantaneous values, though they probably vary.

	Luminosities of \Halpha\ and other tracers of 
star formation tend to be higher in the centers of barred galaxies 
than their unbarred counterparts 
(IR, \cite{deJong84}, \cite{Hawarden86}, \cite{Devereux87}; 
radio, \cite{Hummel81}; 
\Halpha, Ho, Filippenko, \& Sargent 1997b, among others).
The enhanced star formation rates (SFRs) in barred 
nuclei have been attributed to abundant star-forming gas 
accumulated by bars, in an assumption that 
star formation efficiency is not 
systematically different in barred and unbarred nuclei.
It has been also reported that the higher SFR in barred 
nuclei is evident in early-type spirals (S0/a--Sbc) but not so in 
late-type spirals (Sc--Sm), which is attributed to different 
properties of the bars (\cite{Devereux87}; \cite{Dressel88};
\cite{Huang96}; \cite{Ho97b}). 
The SFRs in galactic centers reflect 
time-integrated properties of bar-driven gas transport, 
because a central SFR likely correlates better with the amount of gas 
currently in the galactic center than with the rate of gas inflow 
to that region.
Nevertheless the information provided by SFRs is statistical, and 
one can not tell much about the amount of gas in galactic 
centers because star formation efficiency is hardly constant over 
time or among galaxies.

	Although all the pieces of evidence reviewed above point
to bar-driven transport of gas to galactic centers, 
the evidence from direct observations of 
molecular gas, which is the gas to be transported in the inner regions 
of galaxies, has been very limited; and evidence from statistical 
studies of molecular gas is largely lacking.
The only study based on observations of molecular gas 
in a sample of galaxies is, to our knowledge, 
that by Nishiyama (1995) who measured radial 
distributions of CO in disk galaxies at kpc resolutions.   
Excess gas in barred nuclei was suggested and attributed 
to viscous accretion (due to different rotation curves in barred and 
unbarred systems) and bar-driven accretion, though the statistical test
for the excess was not provided. 
Observations of a broad sample 
of barred and unbarred galaxies are needed in order to clarify 
whether there really is more molecular gas in the centers of barred systems, 
how much gas there is and how the gas is distributed in galactic centers,
and how the gas is related to nuclear star formation and AGN.

		In this paper, we use the NRO-OVRO survey 
of nearby spiral galaxies (\cite{Sakamoto99}) 
to address the issues on bar-driven gas transport and its consequences.
We show that condensations of molecular gas with 
sizes $\lesssim 1$ kpc and masses $10^{8}$--$10^{9}$ \Msol\ 
are frequently seen at the 
centers of spiral galaxies, that the degree of gas concentration to 
the central kpc is higher in barred galaxies than unbarred systems, 
and that the amount of molecular gas in the central kpc does not 
correlate with the optical classification of nuclear activity 
(AGN or star formation).
The higher gas concentration in barred galaxies is not only an 
important addition to the evidence for bar-driven gas transport 
but also a constraint to the rate of mass inflow and the fate of bars.
The lack of correlation between the central gas mass 
and a detectable AGN forces us to reconsider the relation between gas 
and AGN.

\section{The NRO-OVRO Survey \label{s.survey}}
	Our CO(J=1--0) imaging survey was conducted toward the 
central regions of 20 nearby spiral galaxies using 
the millimeter-wave interferometers
at the Nobeyama Radio Observatory (NRO) 
and the Owens Valley Radio Observatory (OVRO).
The principal goal of the survey was to determine the distribution
and kinematics of molecular gas in a sample of nearby spiral galaxies
selected with little bias on their morphologies and activities (such
as starburst and Seyfert nuclei).
It was intended that the survey provided a basic dataset for statistical
studies on gas properties and gas-related phenomena in spiral galaxies. 
A companion paper (\cite{Sakamoto99}, hereafter paper I) gives
details of the design and basic results of the survey; relevant
points are summarized below.

	The 20 galaxies were selected with the 
following four criteria: 
(1) inclination $i < 70\arcdeg$; 
(2) declination $\delta > +5\arcdeg$;
(3) integrated CO intensity 
$\int I_{\rm CO}dV \geq 10$ ${\rm K}(T_{A}^{*}) \kms$ 
in at least one position in the galaxy (usually at the center) 
in the FCRAO Extragalactic CO Survey (\cite{Young95}; \cite{KenneyYoung88});
and (4) no evidence of 
significant perturbation (e.g., merging). 
All galaxies satisfying the above criteria except UGC 2855 were
observed.
No selection was made on the basis of nuclear activity, far-IR luminosity,
and bar properties.
The parameters of the sample galaxies are listed in Table \ref{t.sample}.
The distance of the galaxies 
ranges from 4 to 35 Mpc (\cite{Tully88}), the average being 15 Mpc. 
Their morphological types in the Third Reference Catalogue (RC3; \cite{rc3}) 
are SA:SAB:SB=10:9:1 and a:ab:b:bc:c:cd=1:4:2:9:2:2. 
Their blue-band and far-IR luminosities are
$\log(L_{B}/\Lsol) = 10.35 \pm 0.44$ 
(i.e., $L_{B} \sim L^{*}$, roughly comparable to the Milky Way) and  
$\log(L_{\rm IR}/\Lsol) = 10.19 \pm 0.46$, respectively.
The sample represents nearby ordinary $L^{*}$ galaxies with relatively 
bright CO emission in the central kiloparsecs to enable high 
resolution observations.

--- Table \ref{t.sample} ---

\begin{figure*}[t]
{\hfill	\epsfysize=19cm\epsfbox{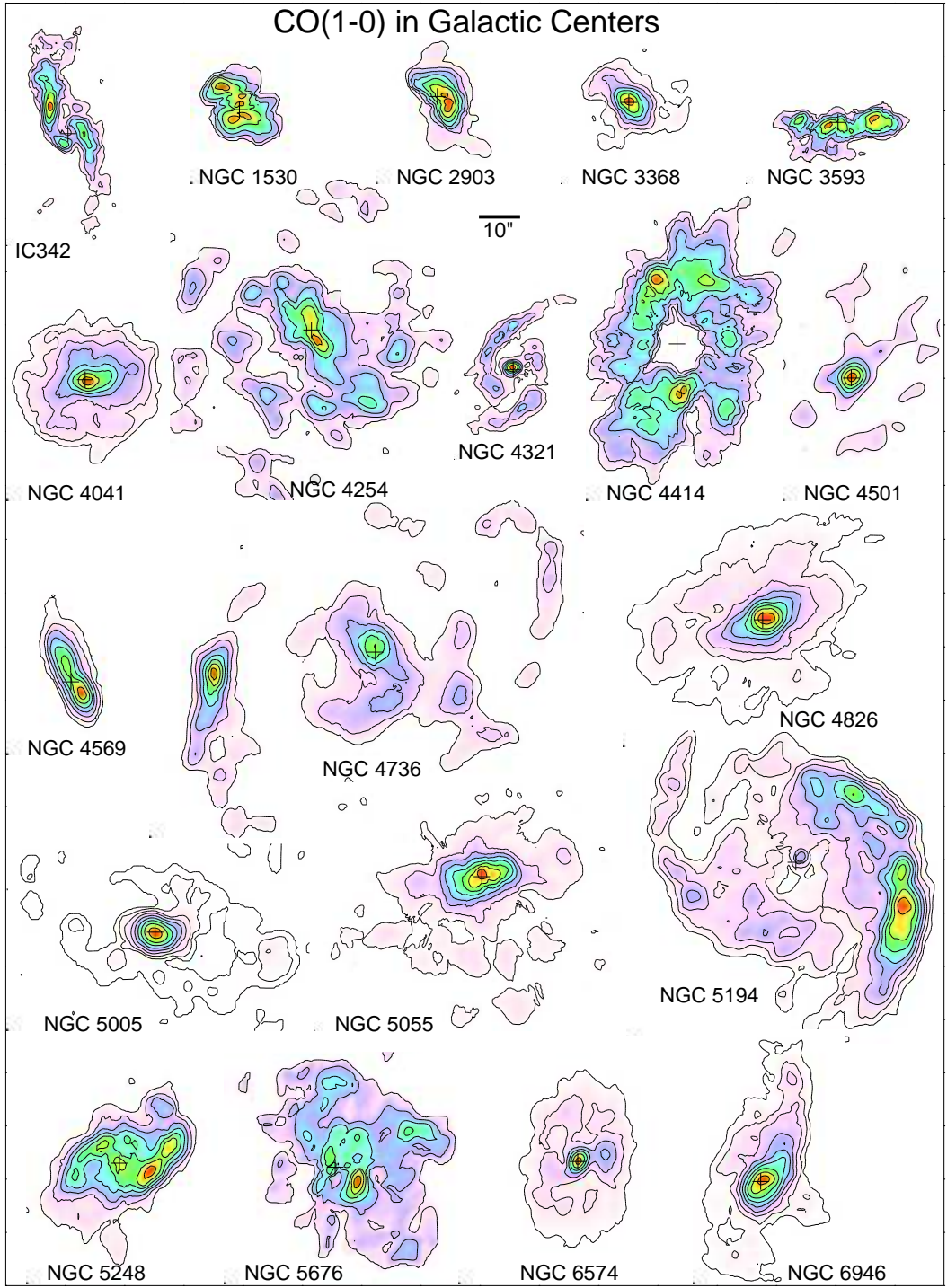}\hfill}
\caption{
CO(J=1--0) in the central regions of 20 spiral galaxies. 
Maps are shown with the same scale on the sky, with
crosses of 300 pc width at the adopted galactic centers. 
No correction for the primary-beam pattern has been applied except 
for NGC 4736, which is a mosaic of three fields. 
Gamma correction, 
in which normalized intensity $I$ ($I=[0,1]$) is mapped to $I^{\gamma}$,
is applied before contouring the maps with various dynamic ranges;
$\gamma=0.6$ for NGC 5005; 
0.8 for NGC 3368, 4041, 4569, 4826, 5194, 6574, and 6946;
1.3 for NGC 4501 and 5676; 
1.0 for the rest.
Contours are at $\frac{1}{2N}$, $\frac{3}{2N}$, ... , and $\frac{2N-1}{2N}$ of
the peak intensity, where $N$ is the number of contours.
The $\gamma$ correction is not applied to the pseudocolour intensities.
\label{f.comaps}}
\end{figure*}

	Our observations have a field of view of 1\arcmin\ (4 kpc
at 15 Mpc), a mean resolution of 4\arcsec\ (300 pc
at 15 Mpc), and velocity resolutions of 5 -- 40 \kms.
The standard sensitivity of the survey was 
40 mJy \beam\ ($1 \sigma$) for a velocity resolution of 10 \kms,
equivalent to $T_{\rm B} = 0.23$ K for a 4\arcsec\ beam. 
The aperture synthesis observations recovered most of CO flux;
the fraction of recovered flux was $70 \pm 14$ \% on average.
The integrated intensity maps are presented in Fig. \ref{f.comaps}.

--- Fig. \ref{f.comaps} ---

\section{Central Gas Condensations}
\subsection{Properties}
	The most apparent feature in the CO maps (Fig. \ref{f.comaps})
is that CO emission peaks toward galactic centers in most galaxies.
The galactic centers marked in the maps are from paper I and
are mostly dynamical centers determined from our CO observations.
The central peaks of CO emission, which we call central gas condensations,
are often very sharp and distinct and sometimes made of twin peaks
or other subfeatures.
The analysis of radial CO distributions in paper I showed that 
CO integrated intensities fall to $1/e$ of the central peaks
at radii less than 500 pc in half of the galaxies, and at
radii less than 1 kpc in 3/4 of the sample.
These scale lengths of the nuclear CO emission are much smaller
than the scale lengths of gas distributions in disks, which are
usually $\gtrsim$ a few kpc.
The sub-kpc scale lengths are not artifacts of aperture synthesis
observations, because we detected most of total flux (see discussion
in paper I).

\begin{figure*}[t]
{\hfill\epsfysize=6cm\epsfbox{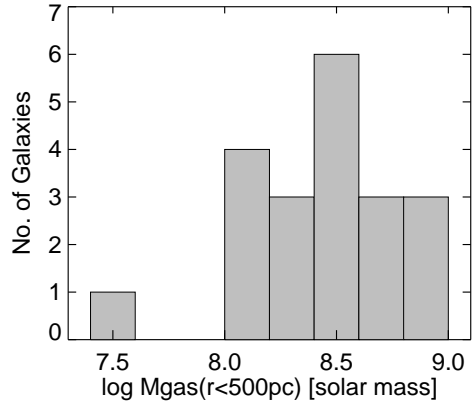}\hfill}
\caption{  
Molecular gas masses within the central kpc derived
from CO emission. 
\label{f.mgascent}}
\end{figure*}

	The mass of molecular gas in the central kpc of
each galaxy is listed in Table \ref{t.sample}, and its histogram is shown
in Fig \ref{f.mgascent}. 
The central gas masses are mostly in the range of $10^{8}$--$10^{9}$ \Msol.
The gas masses are calculated from CO using 
a Galactic CO-to-H$_2$ conversion factor 
$ X_{\rm CO} \equiv N(\HH)/I_{\rm CO} 
	= 3.0\times10^{20}$ cm$^{-2}$ (K \kms)$^{-1}$ 
(\cite{Scoville87}; \cite{Solomon87}; \cite{Bloemen86})
with a correction factor 1.36 for He and heavy elements (\cite{Allen73}). 
The uncertainty of \Xco\ due to the variation of metallicity
is such that \Xco\ is in the range of
1 -- 4 $ \times10^{20}$ cm$^{-2}$ (K \kms)$^{-1}$ (paper I),
which is from the central metallicities $12 + \log({\rm O/H}) = 9.1 \pm 0.2$
and the metallicity--\Xco\ relation proposed by
Wilson (1995) and Arimoto, Sofue, and Tsujimoto (1996).

--- Fig. \ref{f.mgascent} ---

\begin{figure*}[b]
{\hfill\epsfysize=9cm\epsfbox{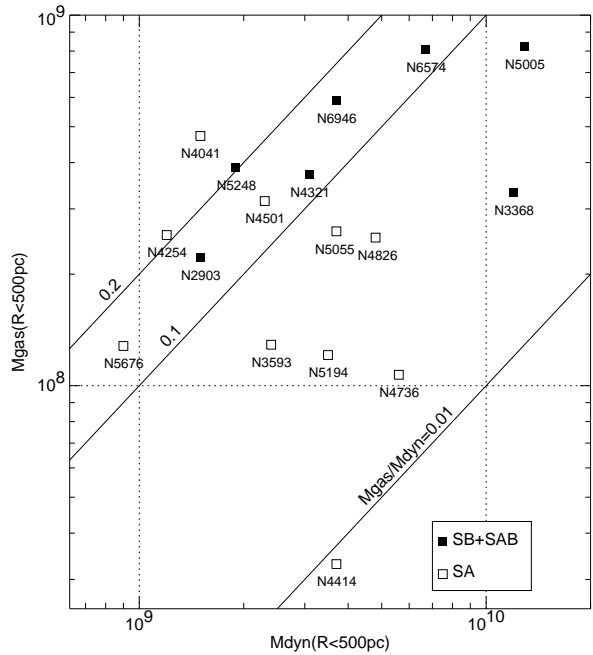}\hfill}
\caption{
 Distribution of molecular gas and dynamical masses within 500 pc 
 from the galactic centers. 
 Barred and unbarred galaxies are plotted with different symbols.
\label{f.mgasmdyn}}
\end{figure*}

	Dynamical masses in the central kpc are calculated 
using the Keplerian formula, $ \Mdyn = R(V/\sin i)^2/G$.
The velocities at $R = 500$ pc are measured from
the CO position-velocity maps in paper I.
The central dynamical masses  
are in the range of $10^{9}$--$10^{10}$ \Msol. 
Fig. \ref{f.mgasmdyn} compares the gas and dynamical masses in the central  kpc.\footnote{
Three galaxies lack \Mdyn\ for the reasons given in Table \ref{t.sample}, 
and hence are absent in the figure.}
The gas-to-dynamical mass ratio is typically 0.1, and 0.3 at maximum.
The error of the Keplerian dynamical mass due to the flatness of mass
distribution is at most 30 \% for an exponential disk (\cite{GalDyn}).
We did not attempt to subtract the effect of noncircular motions, 
because it needs detailed modeling of mass distribution and gas dynamics
in individual galaxies (see, e.g., Wada, Sakamoto, \& Minezaki 1998).
The effects of noncircular motions tend to cancel out
in statistical sense if many barred galaxies are observed
from random viewing angles (see Appendix B of paper I).

--- Fig. \ref{f.mgasmdyn} ---

	Compact CO condensations at galactic centers, such as seen
in this survey, have been seen 
in the Galaxy (Sanders, Solomon, \& Scoville 1984) and
in other nearby galaxies, e.g., NGC 1068  (Planesas, Scoville, \& Myers 1991) 
and four galaxies in Kenney \etal\ (1992), in which three have twin-peak
structure within the condensations.
The scale (size and CO-derived mass) of these CO cores
is similar to that found in our survey.
Taken together, compact CO peaks (or central gas condensations) with
sizes $\sim$ 1 kpc and CO-derived masses $\sim 10^{8}$ \Msol\ appear 
to be prevalent 
in large ($\sim L^{*}$) galaxies, though there are certainly
galaxies without such CO condensations 
(e.g., M31, \cite{Dame93}; M33, \cite{Wilson89}; NGC 4414, \cite{Sakamoto96}).

\subsection{What are they?}
	The  most straightforward interpretation of the central
CO peaks is that they are 
condensations of molecular gas at galactic centers, 
probably caused by dynamical mechanisms such as bar-induced transport of gas. 
Detailed analysis of one of the sample galaxies, NGC 4321, showed that 
it was indeed the case for this moderately barred galaxy 
(\cite{Sakamoto95}; \cite{Wada98}). 
Another possible interpretation of the CO condensations is that 
they are due to enhanced CO emissivity per unit mass of molecular gas, 
presumably caused by extreme physical conditions in galactic centers
such as high pressure causing partly pressure-bound clouds (\cite{Spergel92}).

	It is often argued that the CO-to-\HH\ conversion factor \Xco\ in the
Galactic disk, which we adopt, may overestimate the gas mass in the Galactic
center (e.g., \cite{Sodroski95}; \cite{Oka98}; \cite{Dahmen98}). 
It has not been possible, however, to determine the \Xco\ 
in each galactic center of our sample using the techniques applied
for the Galactic center, and thus there has been no direct evidence 
that CO emissivities are high in the galaxies with CO cores.
We note that the gas-to-dynamical mass ratios in the central
kpc of our sample, $\sim 0.1$, are not very high but
comparable to that in the Galactic disk (\cite{Sanders84}),
which seems to make it unlikely that the gas masses in the
central regions are significantly overestimated with the
Galactic conversion factor.
We also note that the variation of the central gas masses in our sample, 
about an order of magnitude, is unlikely to be dominated
by a random variation of \Xco. 
It is because if the \Xco\ in the galactic centers
have a random scatter of a factor of $\sim 10$ then
there would be no reasonable explanation for the correlation
(discussed in \S \ref{s.concentration})
between bars and the degree of gas (CO) concentration. 
Therefore we interpret the CO condensations to be mostly due to
peaks of molecular gas rather than enhanced CO emissivity, 
though of course some errors in the derived gas masses may exist.

	We also have to consider whether the prevalence of 
the central CO peaks is due to some bias in our sample selection. 
The selection using single-dish CO flux certainly 
precludes galaxies devoid of CO emission in the central kiloparsecs.
However, it is not directly biased toward galaxies with  
sub-kpc CO cores, 
because the single-dish beam (3.3 kpc in FWHM on average) is
much larger than sub-kpc 
and hence the CO emission could be more uniformly distributed 
in the central few kpc. 
On the other hand, it is predicted that self-gravity of molecular gas, 
combined with perturbations from a bar or oval distortion, 
can efficiently drive the gas to the nucleus (\cite{WadaHabe92}). 
Our sample consists of galaxies with larger amount of 
molecular gas in their central regions than average
(provided that CO luminosity traces gas mass).
Thus, it is possible that the fraction of galaxies with central gas
condensations is high in our sample because
of the self-gravitational gas inflow.

\section{Higher Concentration of Molecular Gas in Barred Galaxies
	\label{s.concentration}}
\subsection{Index of gas concentration}
	To quantify the degree of gas concentration in each galaxy we
use the ratio of the surface density of molecular gas 
averaged over the central kpc and that averaged over the
whole optical disk, i.e.,
\begin{equation}
	 f_{\rm con} \equiv \Sgascent/\Sgasdisk .
\end{equation}
The central surface density, \Sgascent, is from our observations
and the disk-averaged surface density, \Sgasdisk, is from 
the total CO flux of the galaxy (Young \etal\ 1995) and 
the galaxy's de Vaucouleurs radius $R_{25}$,
which is the radius at a $B$-band surface brightness of 25 mag arcsec$^{-2}$.
Young \etal\ (1995) observed CO emission along the major axes of the galaxies 
at 45\arcsec\ resolution to at least half of the optical radii $R_{25}$, 
and, if CO emission was detected at multiple positions, extended the 
observations until no detectable emission was seen.
The total CO fluxes were estimated from the radial distributions
with uncertainties of $\sim 25$ \% (\cite{Young95}).
The surface densities and the concentration factors 
are listed in Table \ref{t.sample}.

	The concentration factor \fcon\ is robust against many possible biases
as an index of gas concentration in galactic disks. 
First, it indicates the degree of gas concentration
better than the amount of gas in the central region 
(e.g., $M_{\rm gas}(R < 500 {\rm pc})$)
because \fcon\ is not affected by the total amount of gas in the galaxy.
Second, \fcon\ does not depend on galaxy distance if the distance is correct.
A distance error would affect \Sgascent\ through the size of 
the averaging area but would not affect \Sgasdisk.
Third, the missing flux in our observations does not change \Sgascent\ much,
not only because we detected most of total flux but also
because the extended missing flux must have much lower surface densities 
in the central kpc 
than the detected emission that is peaked at most galactic centers.
Finally and most importantly, the uncertainty in the CO-to-\HH\ 
conversion factor \Xco\ does
not directly affect the concentration factor. 
If \Xco\ is constant in each galaxy, then the constant value can 
be different from one galaxy to another without affecting \fcon,
because the \Xco\ in \Sgascent\ and \Sgasdisk\ cancels out.
This is true in a more general case where \Xco\ varies with
radius but in the same manner, i.e., $X_{\rm CO} = a\times g(R)$ with
$g(R)$ being similar in galaxies.
In this case, \fcon\ correctly indicates the degree of gas concentration
even if the multiplier $a$ is different among galaxies.
A probable cause for the variation of $g(R)$ among galaxies 
is the metallicity--\Xco\ correlation (Wilson 1995; Arimoto \etal\ 1996) 
combined wit the different metallicity gradients in 
barred and unbarred galaxies (see \S 1).
In barred galaxies, the shallower metallicity gradients tend
to decrease \fcon\ than in unbarred systems.
This possible bias, however, is opposite to what we observed (see below),
and thus does not weaken our conclusion.

\begin{figure*}[t]
{\hfill\epsfysize=9cm\epsfbox{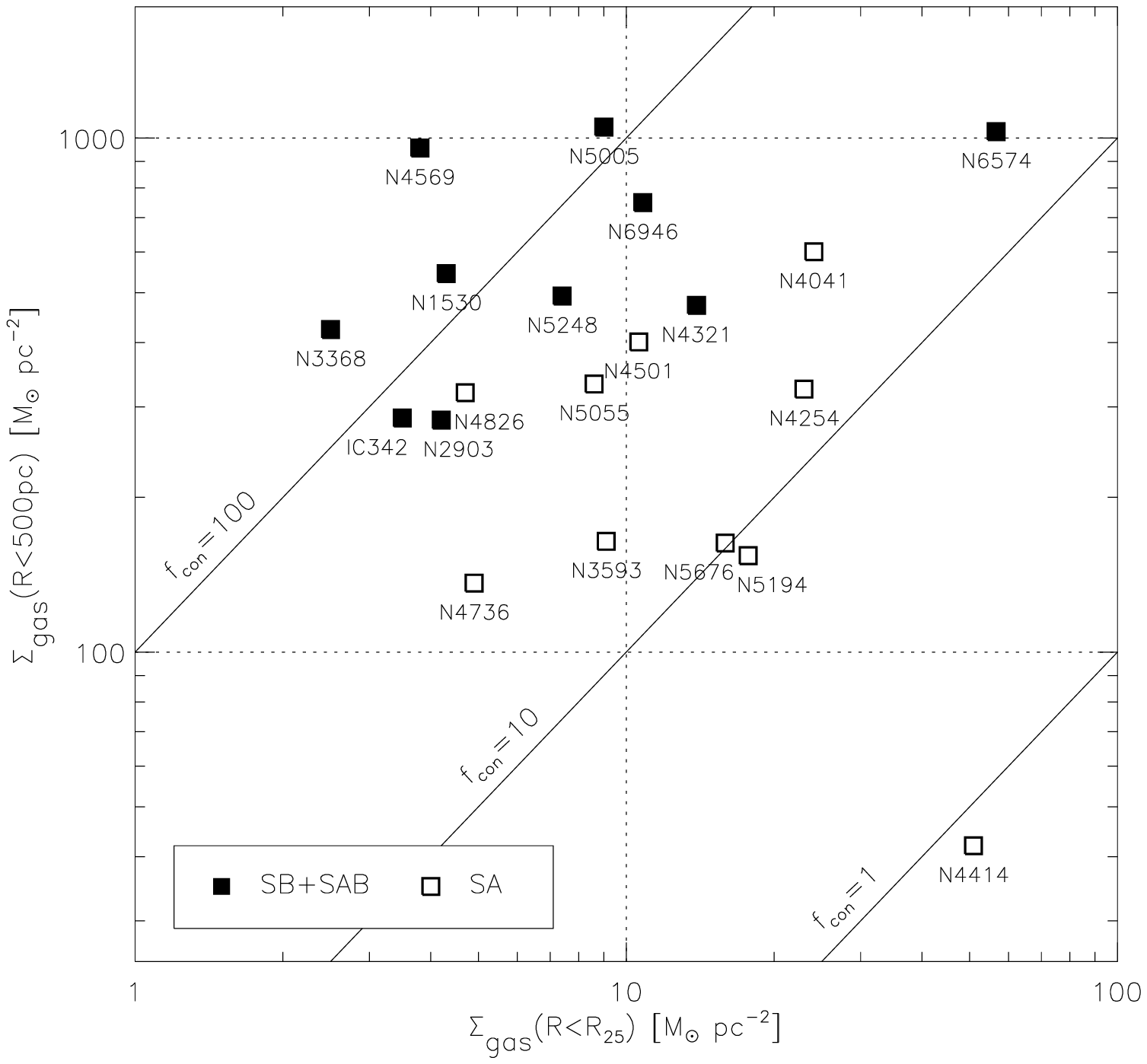}	
 \hspace{0.5cm}		
 \epsfysize=9cm\epsfbox{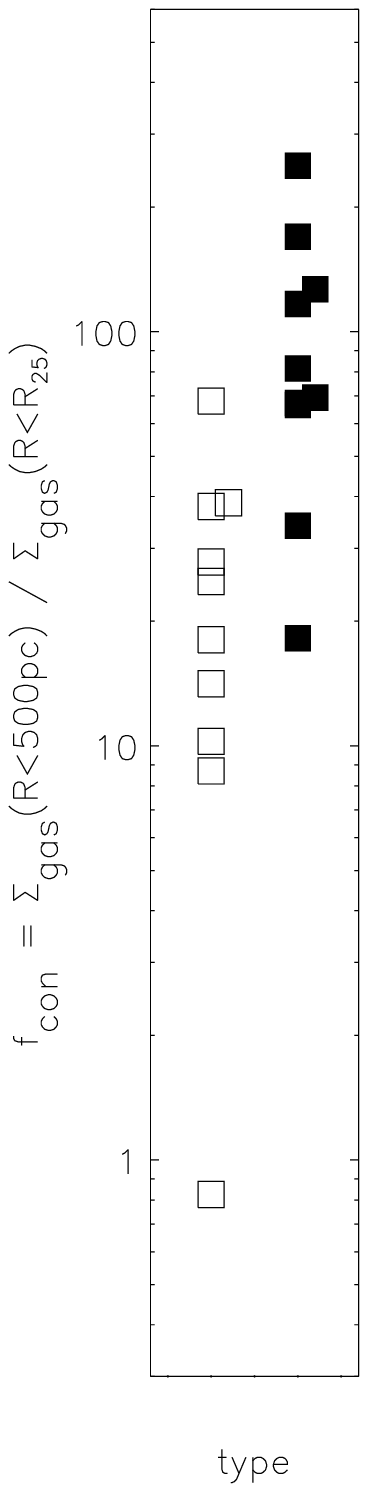}	
\hfill}
\caption{ 
 (Left) Surface densities of molecular gas averaged within the central
 kpc are compared to those averaged over the optical galactic disks.
 The ratio of the central-to-disk averaged surface densities is an
 index of gas central concentration.
 Galaxies in the upper-left part of the panel have higher ratios, i.e.,
 higher gas concentrations.
 (Right) Distribution of the surface density ratio (i.e., concentration
 factor \fcon) for barred galaxies (filled square) 
 and unbarred galaxies (open square).
\label{f.fcon}}
\end{figure*}

	Figure \ref{f.fcon} plots the central and disk-averaged surface
densities of molecular gas and the gas concentration factors with
different symbols for barred and unbarred galaxies.
It clearly shows the higher degree of gas concentration in barred galaxies 
than in unbarred galaxies;
the concentration factor \fcon\ is on average four times higher 
in barred systems.
The Kolmogorov-Smirnov test shows that the difference in \fcon\ 
between barred and unbarred systems is statistically significant; 
the probability for the null hypothesis of no difference between 
the two classes is only $P_{\rm K-S} = 0.007$.

--- Fig. \ref{f.fcon} ---

\begin{figure*}[t]
{\hfill\epsfysize=9cm\epsfbox{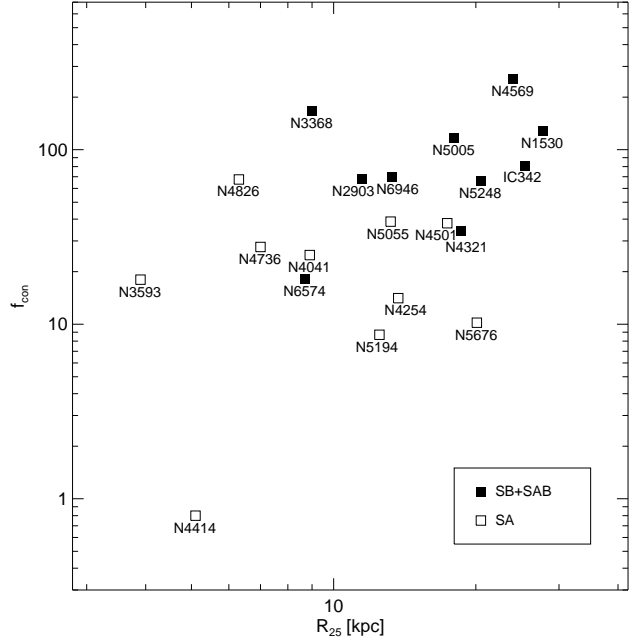}\hfill}
\caption{
 The optical size of galactic disks versus the gas concentration factor \fcon.
 Barred and unbarred galaxies are plotted with different symbols.
\label{f.r25-fcon}}
\end{figure*}

\begin{figure*}[b]
{\hfill\epsfysize=9cm\epsfbox{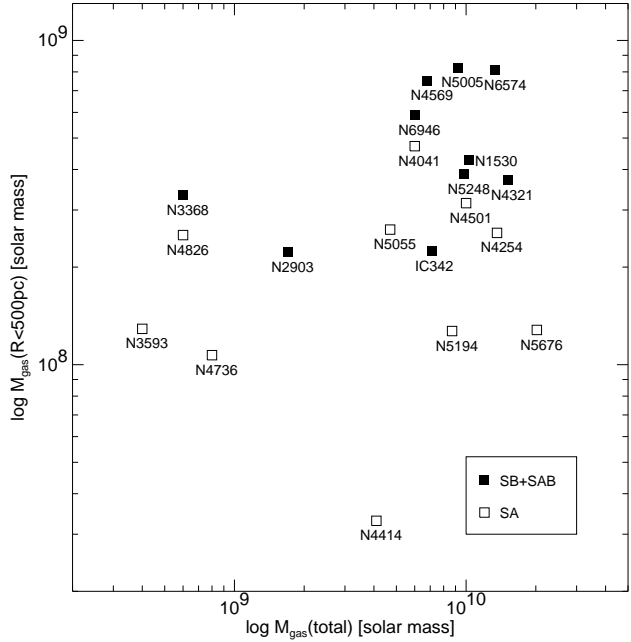}\hfill}
\caption{
 Molecular gas masses within the central kpc are compared
 to those of the entire galaxies.
 Barred and unbarred galaxies are plotted with different symbols.
\label{f.mtot-mcen}}
\end{figure*}

	The concentration factor may be biased by metallicity,  
as we mentioned, and may tend to be smaller in barred galaxies 
than unbarred systems for the same gas distribution.
The correction for this would only {\it increase} the difference 
in \fcon\ between barred and unbarred galaxies.
Fig. \ref{f.r25-fcon} plots \fcon\ versus $R_{25}$.
The barred galaxies in our sample have, on average, larger sizes
than unbarred galaxies.
Larger galaxies have lower \Sgasdisk\ and higher \fcon\ for the same
total amount of gas.
One might think for this reason that the higher \fcon\ in the barred systems 
in our sample is due to their larger sizes.
However, since larger galaxies tend to have larger amount of gas
(in our sample and in general), the bias in \fcon\ due to
galaxy size is small.
In any case, the higher \fcon\ in barred than unbarred systems is
still observed when we set the mean $R_{25}$ of the two classes same
by dropping four largest and four smallest galaxies.
	Figure \ref{f.mtot-mcen} compares the total mass of molecular gas 
in each galaxy with the mass of molecular gas in the central kpc.
The ratio of the two masses is another index of gas concentration,
though it is more susceptible to galaxy size if larger galaxies
have larger total gas mass (i.e., \Sgasdisk\ is roughly constant,
for the same far-IR luminosity; see Casoli \etal\ 1998).
The figure shows that for the same total mass of molecular gas, the
gas masses in the central kpc are higher in barred galaxies
than in unbarred galaxies, being consistent with our finding in 
Fig. \ref{f.fcon}.
We conclude therefore that 
{\it molecular gas is more concentrated to the central kpc
 in barred galaxies than in unbarred galaxies}.

--- Fig. \ref{f.r25-fcon} ---

--- Fig. \ref{f.mtot-mcen} ---

\subsection{Bar-driven gas transport}
	The higher gas concentration in barred galaxies 
strongly supports the bar-driven transport of gas to galactic centers.
The result is the first one that is based on observations of molecular gas,
obtained from a sample of galaxies, and passed a statistical test.
It is therefore complementary to the existing pieces of observational
evidence for the bar-driven gas transport.
In particular, our result supports the conjecture that the higher 
SFRs in barred nuclei is due to larger amount of star forming material there.
Our sample is mostly in Sbc or earlier types and thus
our result could explain the higher SFRs in the nuclei of 
barred early-type spirals.

	The four-fold increase (on average) of the \fcon\ 
in barred galaxies provides an important clue
on the amount of gas transported to the galactic centers.
To increase \fcon\ by a factor of four in a galaxy, 
three times more gas than previously in the central kpc
must be funneled there.
One might think that barred galaxies originally had larger
amount of gas in their centers.
However, this seems less likely than the gas transport after galaxy 
formation, because some barred galaxies must have been formed 
from unbarred galaxies by tidal interactions.
Thus it is most likely that 
{\it more than half of molecular gas within the central
kpc of barred galaxies was transported there from outside by bars}.
If the higher \fcon\ is solely due to the bar-driven gas transport
then the amount of transported gas should be more than $\sim 10^{8}$ \Msol,  
excluding the gas that has been already consumed.

	Finally we note on the distinction between barred and 
unbarred galaxies.
We called SB and SAB galaxies in RC3 as barred and SA galaxies
as unbarred. 
The SB and SAB classes are combined because we have only
one SB galaxy (NGC 1530).
In reality, the barred and unbarred classes are continuous 
rather than discrete.
It is usual to find a weak bar or oval distortion in almost
every disk galaxy when the galaxy is observed in red or near-IR light
(e.g., \cite{Zaritsky86}).
Indeed, there are a few SA galaxies in our sample, e.g., NGC 4254 and 4736,
that exhibit gas distributions probably caused by minor bars (paper I). 
On the other hand, some barred galaxies, e.g.,
NGC 5248 and 6946, may be regarded as unbarred galaxies with a pair of
open spiral arms that mimics a bar (paper I; \cite{Regan95}).
The morphological classification in RC3 is based on
optical imagery of galaxies.
Bars in our conclusion thus refer to the elongated structures 
recognizable (for those who classified the galaxies) in optical 
photographic images.
Such bars must cause larger distortions in gravitational potential
than small bars and weak oval distortions missed in the classification.
Even if a bar really is made of open spiral arms, the elongated
mass distribution causes the oval distortion in gravitational
potential needed to transport gas.
Therefore our result obtained using the optical classification
is in accord with a reasonable conjecture that galaxies with 
stronger oval distortions have larger power to transport gas to their
centers.

\section{Fueling star formation and AGN}
\subsection{Star formation and gas supply}
	The higher concentration of molecular gas in barred galaxies
not only demonstrates the power of stellar bars to fuel
galactic centers but also constrains the relation
between the supply and consumption of star forming gas in the central regions.
The total amount of gas transported so far to the central kpc of
barred galaxies must be larger than the total amount of gas consumed there
mainly by star formation,
because otherwise we would not see the higher \fcon\ in the 
barred nuclei.
Dividing the above amounts of gas and stars by the age of the bar 
in each galaxy one obtains a constraint
that the time-averaged rate of gas inflow must be larger than
the time-averaged SFR in the central kpc, 
i.e., 
\begin{equation}
	\langle	\dot{M}_{\rm bar} \rangle 
	> \langle SFR \rangle,
\label{eq.sfr}
\end{equation}
where $\dot{M}_{\rm bar}$ is the inflow rate of gas 
to the central kpc and the angle brackets denote an average
over the age of the bar.
The ages of bars are likely different among galaxies in our sample.
Therefore the above relation probably hold in a broad range of time
(if not always) in the lifetime of a bar.
The constraint (\ref{eq.sfr}) also gives a crude estimate of the time-averaged
rate of gas inflow if one substitutes the time-averaged SFR 
with an ensemble average of current SFRs.

	We estimate the SFR in the central kpc of each galaxy in two ways, 
from \Halpha\ and from far-IR.
Derived SFRs are in Table \ref{t.halpha}.
The SFRs from \Halpha\ and FIR agree well 
(except for NGC 5194) considering their uncertainties 
(at least a factor of a few for each). 
The \Halpha\ data are from the $2'' \times 4''$ aperture
observations of Ho \etal\ (1997a).  
In each galaxy, we estimate \Halpha\ luminosity of the central kpc
assuming that the line intensity is uniform in the area, 
then correct it for extinction using 
the \Halpha\ to \Hbeta\ ratio (Ho \etal\ 1997a),
and calculate the SFR using the coefficient in
Kennicutt, Tamblyn, \& Congdon (1994).
The SFRs are a factor of two lower (on average) if
the \Halpha\ luminosities are estimated from the larger ($8''$) 
aperture observations by Keel (1983).
The far-IR data are from the IRAS survey, which are tabulated in
the Table 2 of paper I along with the total FIR luminosities.
The FIR luminosity from the central kpc is estimated
using the 
$S_{10 \micron}(r \le 0.5 \rm kpc)/S_{10 \micron}^{\rm total}$
ratio.
We compute $S_{10 \micron}(r \le 0.5 \rm kpc)$
from ground-based observations (of 5\arcsec\ -- 20 \arcsec\ apertures)
assuming the uniformity of 10 \micron\ intensity in the central kpc,
and $S_{10 \micron}^{\rm total}$ by extrapolation from
the IRAS 12 \micron\ data with a color correction (\cite{Devereux87}).
The conversion from \Lfir\ to SFR uses the mean of
the coefficients derived by Kennicutt (1998) and Buat \& Xu (1996)
for starbursts and more quiescent spirals (Sb and later), respectively.

	The star formation rates in the central kpc of barred
galaxies in our sample are mostly in the range of 0.1 -- 1 \Msolyr.
As we saw above, it gives a lower limit to the time-averaged rate of
gas inflow $\langle \dot{M}_{\rm bar} \rangle$.
The value is consistent with theoretical predictions of 
$\dot{M}_{\rm bar} \sim$ 0.1--10 \Msolyr\
and the instantaneous $\dot{M}_{\rm bar}$ estimated in a few barred galaxies
(see \S \ref{s.intro}).
Note that, however, the equation (\ref{eq.sfr}) does not  
require the current $\dot{M}_{\rm bar}$ to be larger
than the current SFR.
It is possible that the mass inflow rate has been declining
because the amount of gas available from the outer regions is limited, 
and that the SFR has been increasing as the
star forming gas in the central region has been increasing.
Thus the current $\dot{M}_{\rm bar}$ could be below the current SFR
and also below the lower limit of $\langle \dot{M}_{\rm bar} \rangle$. 

--- Table \ref{t.halpha} ---

\subsection{AGN and gas condensations}
\begin{figure*}[t]
{\hfill\epsfysize=9cm\epsfbox{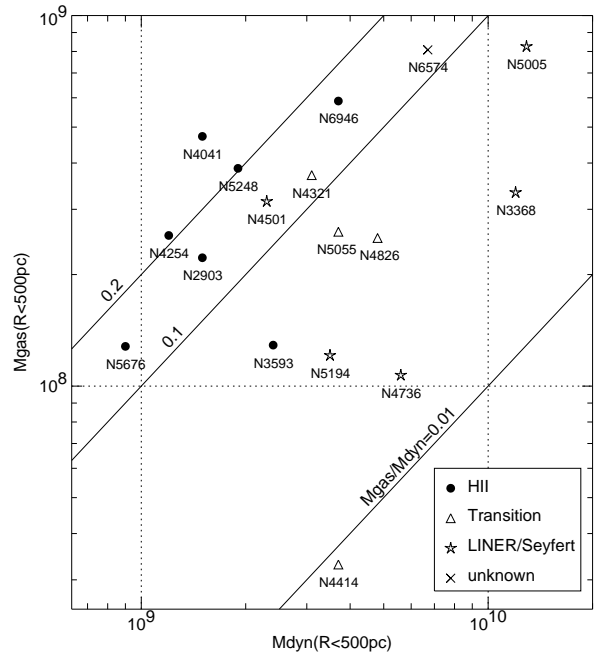}\hfill}
\caption{
 Gas and dynamical masses within the central kpc of galaxies.
 The data are the same as in Fig. \ref{f.mgasmdyn} 
 but are plotted with symbols representing  
 the classification of the nuclear optical spectrum; 
 HII, transition (i.e., HII + LINER), and Seyfert or LINER.
\label{f.q-type}}
\end{figure*}

	Figure \ref{f.q-type} plots the gas and dynamical
masses in the central kpc using symbols 
according to the types of nuclear activity, which are 
from optical spectroscopy by Ho \etal\ (1997a).
By looking only the gas masses, the vertical axis of the plot, 
it is apparent
that the amount of gas in the central kpc does not determine
the type of nuclear activity;
the gas masses are not different for HII, 
transition type (= intermediate type between LINER and HII), 
and LINER or Seyfert nuclei.
In particular, we note that the large amounts of gas concentrated 
within 500 pc are not necessarily accompanied by AGN, which
is in Seyferts and probably in LINERs and transition nuclei. 
A similar result has been obtained by Vila-Vilaro \etal\ (1999) who
made a CO survey of Seyfert and non-Seyfert galaxies
(Vila-Vilaro, Taniguchi, \& Nakai 1998) and did not find a
significant difference in the amount of molecular gas in their 
16\arcsec\ ($\sim 1$ kpc) beam.
These results suggest that the gas fueling to this scale ($R \sim 500$ pc)
does not determine the nature of the nuclear activity.
This could explain why no correlation is found between bars and Seyfert
activity (\cite{Ho97b}; \cite{Mulchaey97}) despite the higher
concentration of molecular gas in barred nuclei 
as we saw in \S \ref{s.concentration}.

--- Fig. \ref{f.q-type} ---

	Figure \ref{f.q-type} also shows a segregation between HII nuclei 
and active nuclei (i.e., Seyfert, LINER, and transition)
according to the gas-to-dynamical mass ratio.
The gas-to-dynamical mass ratios in HII nuclei are significantly larger
than those in active nuclei\footnote{
 We could not measure \Mgas/\Mdyn\ in three galaxies from our data.
 For IC 342, however, Sage \& Solomon (1991) obtained 
 $M_{\rm H_{2}}/\Mdyn = 0.27$ within $r = 1.5$ kpc.
 If corrected for the distance of 3.9 Mpc adopted by us, 
 or for $D = 1.8$ Mpc adopted by Turner \& Hurt (1992),
 \Mgas/\Mdyn\ would be 0.32 within 1.3 kpc and
 0.15 within 0.6 kpc, respectively. 
 ($\Mgas = 1.36 M_{\rm H_{2}}$.)
 The gas-to-dynamical mass ratio inside 0.5 kpc is probably larger than the
 above values, because of the gas central concentration in IC 342.
 Even the smallest value of \Mgas/\Mdyn\ = 0.15 is larger
 than the apparent boundary of $\approx 0.1$ between HII and AGN
 classes, being consistent with the HII type of IC 342.
 Thus it would just strengthen our observation if IC 342
 could be plotted in Fig. \ref{f.q-type}.  
}, 
judging from the Kolmogorov-Smirnov test (the probability for no
difference is $P_{\rm K-S} = 0.002$).
The larger gas-to-dynamical mass ratios 
(and hence smaller $Q$ values [see below]) in non-AGN galaxies
than in galaxies with active nuclei have also been noticed by
Vila-Vilalo \etal\ (1999) and Kohno \etal\ (1999).

	A straightforward interpretation for this is that galaxies 
with larger 
amounts of gas for their sizes have more active star formation, 
resulting in nuclear spectra dominated by HII regions. 
A simple assessment of the stability of the nuclear gas disk 
supports this idea. 
If a uniform gas disk with a surface density $\Sigma_{\rm gas}$ 
has a rotational velocity $V$ at a radius $R$, 
its gas-to-dynamical mass ratio within $R$ is
$ M_{\rm gas}/M_{\rm dyn} = \pi G R \Sigma_{\rm gas}/V^2 $.
The stability parameter $Q$ of the same disk 
(\cite{Safronov60}; \cite{Toomre64}) is
$ Q = \kappa\sigma / \pi G \Sigma_{\rm gas} \approx 2V\sigma / \pi G R \Sigma_{\rm gas} $
where $\kappa $ and $\sigma $ are epicycle frequency and gas 
velocity dispersion, respectively.
We have used the fact that the rotation curves are steeply rising 
in the nuclear regions (paper I) and hence 
the epicyclic frequency $\kappa $ is $\kappa \approx 2V/R$. 
The two parameters have a relation
\begin{equation}
    Q \approx 2 \frac{\sigma}{V} \left(\frac{M_{\rm gas}}{M_{\rm dyn}}\right)^{-1}.
\end{equation}
Thus the gas-to-dynamical mass ratio is closely related to the $Q$ parameter. 
Numerically, $Q$ is unity for typical values of $\sigma = 10$ \kms\ 
and $V = 200$ \kms\ when $M_{\rm gas}/M_{\rm dyn}$ is 0.1. 
The gas disk is unstable when $Q$ is below unity, and the disk instability 
likely results in cloud formation and subsequent star formation. 
Therefore it is not surprising that HII nuclei appear predominantly 
above the line of $\Mgas / \Mdyn \approx 0.1$. 

	The average H$\alpha$ luminosity of the HII nuclei is 
larger than that of the active nuclei (see Table \ref{t.average}), 
though the sample is very small. 
Therefore it is possible that emission from a dwarf AGN, 
which may exist in most spiral galaxies, 
is overwhelmed by the emission from nuclear HII regions 
when the nuclear gas disk is unstable, 
resulting in the HII spectral classification for the galaxy.
Such hidden AGN in HII nuclei have been inferred from models
of diagnostic line ratios with more than one type of
ionizing source, and from optical spectroscopy
at high-spatial resolutions (e.g., Kennicutt, Keel, \& Blaha 1989). 
Similar situations may be prevalent.

--- Table \ref{t.average} ---

	There are of course caveats for the above argument on disk 
stability, and there exist other possible causes that could
explain the lack of correlation between the central gas mass and AGN.
The errors in the $Q$ values include 
the uncertainty of \Xco\ and likely internal structure in the CO cores.
The alternative causes for the apparent independence of gas mass and AGN
include the difficulty of delivering gas
from $R \sim 500$ pc to the central accretion disk,
the small ($\sim 100$ pc) central holes of gas that are inferred
in several galaxies (paper I), 
and the higher chance of obscuration for the AGNs in gas-rich nuclei.
Probably all of these, including the effect of disk stability, 
contribute to the relation between the central
gas mass and the appearance of emission line spectrum.

\section{Constraints on Secular Evolution \label{s.evolution}}
	Theoretically it is predicted that the bar-driven
transport of significant amount of gas to the center of a galaxy
eventually destroys the bar 
(\cite{Hasan90}; \cite{Pfenniger90}; \cite{FriedliBenz93}).
The critical mass for the bar dissolution has been estimated to be
about 5 \% of the total mass of disk and bulge for the model of 
Norman, Selwood, \& Hasan (1996).
For the Galaxy, which is comparable to our $\sim L^{*}$ sample,
the total mass of disk and bulge is about
$5 \times 10^{10}$ \Msol\ (\cite{Dehnen98}), 
and the critical mass is a few $10^{9}$ \Msol.
The timescale for the bar dissolution is very short, 
being comparable to the dynamical timescale of the bar or
a few $10^{8}$ yr (\cite{Norman96}). 
However, as noted by Norman \etal\ (1996), there are many
parameters involved in the bar dissolution, and they may well change the
critical mass and the dissolution timescale.
For example, the following parameters have not been fully explored;
the shape of the bar, that of the disk,
and the spatial distribution of funneled gas in the central regions
(some simulations use a central point source with a variable mass
 instead of a gas condensation with finite extent). 
In a different model, a bar was destroyed by a smaller amount 
of mass concentration, 0.5 \% of the disk mass or
a few $10^{8}$ \Msol\ for the Galaxy,
at a longer timescale of a few $10^{9}$ yr (\cite{Hozumi98}).
Berentzen \etal\ (1998) also obtained a longer timescale for bar dissolution,
$\sim 2\times 10^{9}$ yr or $\sim 7$ rotations, in their
(stars + gas in a halo) model of a galaxy.

	Our observations can constrain the timescale
as well as the critical mass of the bar dissolution.
First, our sample galaxies are among galaxies with largest amount
of molecular gas in their inner regions ($\sim 4$ kpc) and hence
their gas masses in the galactic centers, 
$M_{\rm gas}(R < 500 {\rm pc}) \sim 10^{8}$ -- $10^{9}$ \Msol,
are probably among the largest in ordinary spiral galaxies.
Thus if these gas condensations (plus the amount of stars formed from
the gas) are not enough to destroy bars,
then the bar-dissolution would be rare in galaxies of
similar scales ($L \sim L^{*}$).
Second, if a bar can be destroyed by a gas condensation 
of $\sim 10^{8}$ -- $10^{9}$ \Msol, then the different degree of gas
concentration between barred and unbarred galaxies constrains
the lifetime of the bars.
It is because the bars responsible for the higher gas concentrations
must be alive until a large part of the funneled gas is consumed;
otherwise we would frequently observe currently unbarred 
(but previously barred) galaxies with high degree of gas 
concentration due to the destroyed bars.
This argument leads to a constraint on timescales for
bar dissolution and gas consumption,
\begin{equation}
	\tau_{\rm bar\;dissolution} \geq \tau_{\rm gas\;consumption}. 
\end{equation}
The gas consumption time may be evaluated using the current SFR and
the current mass of molecular gas in the central region.
\begin{equation}
	\tau_{\rm gas\;consumption} \gtrsim \frac{M_{\rm gas}^{\rm center}}{SFR}
\end{equation}
The inequality is due to the ignored return of gas from stars
and the supply of gas by a bar, 
though the latter might be small in old barred galaxies because of
the limited supply of gas from the outer regions.
At the current SFR of 0.1 -- 1 \Msolyr\ the consumption
times of the central gas condensations of $\sim 10^{8}$ -- $10^{9}$ \Msol\
are larger than $10^{8}$ -- $10^{10}$ yr.
The smallest value does not contradict with the shortest timescale
predicted by simulations.
On the other hand, if larger values are the case in many spiral galaxies, 
then the bar dissolution would take longer time than predicted
or would not be caused by a gas concentration of 
$\sim 10^{8}$ -- $10^{9}$ \Msol.
The latter case does not contradict with the critical mass
estimated by Norman \etal\ (1996), but makes the bar
dissolution in $L^{*}$ galaxies a rare phenomenon caused only by
exceptionally large transport of gas.

	The above argument on the timescales for bar dissolution
and gas consumption is an exploratory one, 
and may involve considerable errors for individual galaxies
because, for example, 
the gas consumption may be much faster owing to
episodic starbursts.
However, statistical analysis of this kind is one of a few ways to
observationally constrain the evolution of bars, and thus worthwhile
to pursue.
Also, the measurement of the gas concentration is a way to find
out galaxies with young bars and galaxies whose bars have been 
recently destroyed.
The former would have low \fcon\ for barred galaxies, while
the latter would have high \fcon\ ($\gtrsim 100$) 
for unbarred galaxies.
There probably are galaxies with young bars in interacting systems 
(\cite{Noguchi88}).
The galaxies in which bars have recently been destroyed would be 
most interesting to discover in order to elucidate the
conditions and mechanisms for bar dissolution.

\section{Conclusions}
	From our CO(1--0) imaging survey of the central regions of 
20 nearby spiral galaxies, we find:

1.  Strong peaks of CO emission, with radial scale lengths $\lesssim 500$ pc
and CO-derived masses $\sim 10^{8}$--$10^{9}$ \Msol\ in the central kpc,
are prevalent in the nuclei of gas-rich $\sim L^{*}$ spiral galaxies.
These central gas condensations constitute about 
10\% of the dynamical masses in the central kpc.

2. Molecular gas is more concentrated to the central kpc
in barred galaxies than in unbarred systems.
This strongly supports the theory of bar-driven gas transport
to galactic centers.
More than half of molecular gas within the central kpc
of barred galaxies was transported there from outside by bars.

3. The higher gas concentration in barred galaxies
suggests that the statistically higher SFRs in barred nuclei
are due to abundant star-forming material there,
and that star formation in the nuclear regions has not been
able to catch up with the supply of gas by bars.
A lower limit to the time-averaged rate of mass inflow to the central kpc
is 0.1 -- 1 \Msolyr.

4. No correlation is found between nuclear activity (AGN) and the 
mass of molecular gas within 500 pc of the nucleus.
The similar incidence of AGN between barred and unbarred galaxies 
is thus not due to the problem of bars to transport gas to the central
kpc but must be due to the difficulties in gas fueling within 
the central kpc and in detecting the dwarf AGN.

5. The gas-to-dynamical mass ratio in the central kpc is 
higher in galaxies with HII spectral classification than in galaxies 
having an AGN, with the dividing line at $\Mgas / \Mdyn \approx 0.1$.
It may well be related to the stability of the nuclear 
molecular gas disk, which is lower when $\Mgas / \Mdyn$ is high. 
In HII galaxies, star formation triggered by the instability may be 
overwhelming dwarf active nuclei.

6. The different degrees of gas concentration in barred 
and unbarred galaxies would not have been observed if gas condensations
of $10^{8}$ -- $10^{9}$ \Msol\ in the central kpc 
could destroy bars on a timescale shorter than the gas consumption times.
The gas consumption time, or a lower limit to the
bar dissolution timescale, is in the range of $10^{8}$ -- $10^{10}$ yr.
The lowest value does not contradict models of bar dissolution
while the highest value implies that bar dissolution is not caused by
the $10^{8}$ -- $10^{9}$ \Msol\ gas condensations in $\sim L^{*}$
galaxies.
Measurement of gas concentration in more galaxies would provide
an observational test on the bar dissolution scenario.

\acknowledgements
	We are grateful to NRO and OVRO for the generous allotments of 
observing time, and to the observatory staffs whose work on the arrays 
enabled our survey. Discussions with colleagues at NRO and Caltech/OVRO,
in particular comments from Andrew Baker to an early draft, 
greatly helped to clarify our points.
We also thank the referee, Dr. Jean Turner, for her 
valuable comments and suggestions.
K.S. is supported by Grant-in-Aid for JSPS Fellows. 
OVRO is funded by NSF grant AST 96-13717.

\clearpage

\begin{deluxetable}{llcrrcccccr}
\tablecaption{Parameters of individual galactic nucleus \label{t.sample}}
\tablewidth{0pt}
\tablehead{
 \colhead{name} & 
 \colhead{morph. type\tablenotemark{a}} &
 \colhead{class\tablenotemark{b}} &
 \colhead{$D$\tablenotemark{c}} &
 \colhead{$R_{25}$\tablenotemark{d}} &
 \colhead{$M_{\rm gas}^{\rm 500\,}$\tablenotemark{e}} &
 \colhead{$M_{\rm dyn}^{\rm 500\,}$\tablenotemark{f}} & 
 \colhead{$f_{\rm gas}^{\rm 500\,}$\tablenotemark{g}} & 
 \colhead{$\Sigma_{\rm gas}^{500\,}$\tablenotemark{h}} &
 \colhead{$\Sigma_{\rm gas}^{\rm disk\,}$\tablenotemark{i}} &
 \colhead{\fcon\tablenotemark{j}}    \nl
 \colhead{ } & 
 \colhead{ } &
 \colhead{ } &
 \colhead{Mpc} &
 \colhead{kpc} &
 \colhead{$10^7 M_{\odot}$} & 
 \colhead{$10^9 M_{\odot}$} & 
 \colhead{\%} &
 \colhead{$M_{\odot}$pc$^{-2}$} & 
 \colhead{$M_{\odot}$pc$^{-2}$} & 
 \colhead{\%} 
}
\startdata
  IC 342 & SAB(rs)cd   &    H &  3.9 & 25.4  &	22.4 &  \nd &  \nd &   285 &  3.5 &  81.1 \nl 
NGC 1530 & SB(rs)b     &  \nd & 36.6 & 27.7  &	42.8 &  \nd &  \nd &   545 &  4.3 & 127.6 \nl 
NGC 2903 & SAB(rs)bc   &    H &  6.3 & 11.5  &	22.2 &  1.5 & 14.8 &   283 &  4.2 &  68.1 \nl 
NGC 3368 & SAB(rs)ab   &   L2 &  8.1 &  9.0  &	33.3 & 12.0 &  2.8 &   424 &  2.5 & 166.9 \nl 
NGC 3593 & SA(s)0/a    &    H &  5.5 &  3.9  &	12.9 &  2.4 &  5.4 &   164 &  9.1 &  18.0 \nl 
NGC 4041 & SA(rs)bc    &    H & 22.7 &  8.9  &	47.2 &  1.5 & 31.5 &   601 & 24.1 &  24.9 \nl 
NGC 4254 & SA(s)c      &    H & 16.8 & 13.7  &  25.5 &  1.2 & 21.2 &   325 & 23.0 &  14.1 \nl 
NGC 4321 & SAB(s)bc    &   T2 & 16.8 & 18.6  & 	37.1 &  3.1 & 12.0 &   472 & 13.9 &  34.0 \nl 
NGC 4414 & SA(rs)c     &  T2: &  9.7 &  5.1  &	 3.3 &  3.7 &  0.9 &    42 & 50.9 &   0.8 \nl 
NGC 4501 & SA(rs)b     &   S2 & 16.8 & 17.4  &	31.5 &  2.3 & 13.7 &   401 & 10.6 &  37.9 \nl 
NGC 4569 & SAB(rs)ab   &   T2 & 16.8 & 24.0  &	75.1 &  \nd &  \nd &   956 &  3.8 & 254.5 \nl 
NGC 4736 & (R)SA(r)ab  &   L2 &  4.3 &  7.0  &	10.7 &  5.6 &  1.9 &   136 &  4.9 &  27.7 \nl 
NGC 4826 & (R)SA(rs)ab &   T2 &  4.1 &  6.3  &	25.1 &  4.8 &  5.2 &   320 &  4.7 &  67.5 \nl 
NGC 5005 & SAB(rs)bc   & L1.9 & 21.3 & 18.0  &	82.5 & 12.9 &  6.4 &  1050 &  9.0 & 116.6 \nl 
NGC 5055 & SA(rs)bc    &   T2 &  7.2 & 13.2  &	26.1 &  3.7 &  7.1 &   332 &  8.6 &  38.7 \nl 
NGC 5194 & SA(s)bc     &   S2 &  7.7 & 12.5  &	12.1 &  3.5 &  3.5 &   154 & 17.7 &   8.7 \nl 
NGC 5248 & SAB(rs)bc   &    H & 22.7 & 20.5  &	38.7 &  1.9 & 20.4 &   493 &  7.4 &  66.2 \nl 
NGC 5676 & SA(rs)bc    &    H & 34.5 & 20.1  &	12.8 &  0.9 & 14.1 &   163 & 15.9 &  10.2 \nl 
NGC 6574 & SAB(rs)bc   &  \nd & 35.0 &  8.7  &	80.8 &  6.7 & 12.1 &  1029 & 56.6 &  18.2 \nl 
NGC 6946 & SAB(rs)cd   &    H &  5.5 & 13.3  & 	58.8 &  3.7 & 15.9 &   749 & 10.8 &  69.3 \nl 
\enddata
\tablenotetext{a}{Morphological classification in the RC3 (de Vaucouleurs \etal\ 1991).}
\tablenotetext{b}{Spectral classification of the nucleus taken from Ho \etal\ (1997a).
H = HII nucleus, S = Seyfert nucleus, L = LINER, T = Transition object (= HII + LINER). 
The number designates the type (e.g., S2 is a type 2 Seyfert nucleus.)}
\tablenotetext{c}{Galaxy distance derived in Tully (1988) using $H_{0} = 75 $ \kms Mpc$^{-1}$
with correction for the Virgocentric inflow.}
\tablenotetext{d}{Isophotal radius at 25 mag arcsec$^{-2}$ in $B$-band, 
which is corrected for extinction and inclination 
and from RC3 (de Vaucouleurs \etal\ 1991).}
\tablenotetext{e}{Gas mass within 500 pc from the nucleus.}
\tablenotetext{f}{Dynamical mass in 500 pc of the nucleus.
Three galaxies do not have $M_{\rm dyn}$ for the following reasons: 
(IC342) no gas to measure velocity is at a radius of 500 pc on the major axis;
(NGC 1530) spatial resolution is insufficient to determine the rotation curve;
and (NGC 4569) very large noncircular motion in the p-v map 
in Sakamoto \etal\ (1999).
}
\tablenotetext{g}{Gas-to-dynamical mass ratio within 500 pc of the nucleus.}
\tablenotetext{h}{Mean surface density of molecular gas averaged within 500 pc from the nucleus.}
\tablenotetext{i}{Surface density of molecular gas averaged over the galactic disk, 
which is derived from the total CO flux (Young \etal\ 1995) 
and the optical size of the galaxy $R_{25}$.}
\tablenotetext{j}{Concentration factor of molecular gas,
$f_{\rm con} \equiv \Sigma_{\rm gas}^{500}/\Sigma_{\rm gas}^{\rm disk}$.}
\end{deluxetable}

\begin{deluxetable}{llcccccc}
\tablecaption{Star formation rates in galactic centers \label{t.halpha}}
\tablewidth{0pt}
\tablehead{
 \colhead{name} & 
 \multicolumn{2}{c}{SFR($r<0.5$ kpc)} &
 \colhead{ref.} \nl
 \colhead{ } & 
 \multicolumn{2}{c}{\Msolyr} &
 \colhead{ } \nl 
 \colhead{(1)} & 
 \colhead{(2)} &
 \colhead{(3)} & 
 \colhead{(4)} 
}
\startdata
  IC 342  &  0.87  & 0.53 & 1   \nl  
NGC 1530  &  \nd   & \nd  & \nd \nl  
NGC 2903  &  1.12  & 0.48 & 2   \nl  
NGC 3368  &  0.44  & 0.11 & 3   \nl  
NGC 3593  &  0.26  & \nd  & \nd \nl  
NGC 4041  &  0.78  & \nd  & \nd \nl  
NGC 4254  &  0.28  & \nd  & \nd \nl  
NGC 4321  &  0.24  & 0.14 & 4   \nl  
NGC 4414  &  0.02\tm{a} & \nd  & \nd \nl  
NGC 4501  &  0.08  & 0.03 & 4   \nl  
NGC 4569  &  1.26  & 0.22 & 4   \nl  
NGC 4736  &  0.09  & 0.38 & 3   \nl  
NGC 4826  &  0.56  & 0.34 & 3   \nl  
NGC 5005  &  0.10\tm{a} & 0.29 & 4 \nl  
NGC 5055  &  0.12  & \nd  & \nd \nl  
NGC 5194  &  3.46  & 0.06 & 5   \nl  
NGC 5248  &  0.76  & \nd  & \nd \nl  
NGC 5676  &  0.02  & \nd  & \nd \nl  
NGC 6574  &  \nd   & \nd  & \nd \nl  
NGC 6946  &  2.95  & 0.56 & 6   \nl  
\enddata
\tablecomments{
Col. (1): Galaxy name. 
Col. (2): SFR in the central kpc estimated from \Halpha\ (see text).
          $SFR$ [\Msolyr] 
		= $\log L(\Halpha) / 1.24\times10^{41} $ [erg s$^{-1}$]
          (Kennicutt \etal\ 1994).
	  Extinction is corrected using the \Halpha/H$\beta$ ratio 
          with the assumed intrinsic ratio of 2.86.
Col. (3): SFR in the central kpc estimated from infrared data (see text).
	$SFR$ [\Msolyr] = $2.4\times 10^{-10} \log L_{\rm 8-1000 \micron}$ [\Lsol].
Col. (4): References for 10 \micron\ data. 
	1. Becklin \etal\ (1980); 
	2. Wynn-Williams \& Becklin (1985);
	3. Cizdziel, Wynn-Williams, \& Becklin (1985);
	4. Devereux, Becklin, \& Scoville (1987);
	5. Telesco, Decher, \& Gatley (1986);
	6. Rieke (1976) 
}
\tablenotetext{a}{Correction for extinction is not made.} 
\end{deluxetable}

\begin{deluxetable}{lccccc}
\tablecaption{Average parameters in the galactic centers \label{t.average}}
\tablewidth{0pt}
\tablehead{
 \colhead{class} & 
 \colhead{$M_{\rm gas}$($r<500$pc)} &
 \colhead{$M_{\rm dyn}$($r<500$pc)} & 
 \colhead{$M_{\rm gas}/M_{\rm dyn}$} & 
 \colhead{\fcon} & 
 \colhead{SFR($r<500$pc)\tablenotemark{a}}   \nl
 \colhead{ } & 
 \colhead{$10^8 M_{\odot}$} & 
 \colhead{$10^9 M_{\odot}$} & 
 \colhead{\%} &
 \colhead{ } & 
 \colhead{\Msolyr} 
}
\startdata
SB+SAB        & $4.9\pm2.3$ (10) & $6.0\pm4.7$ ( 7) & $12.1\pm5.9$ ( 7) & $100.2\pm69.8$ (10) & $0.97\pm0.84$ (8) \nl
SA            & $2.1\pm1.2$ (10) & $3.0\pm1.5$ (10) & $10.5\pm9.3$ (10) & $24.9\pm18.5$ (10)  & $0.25\pm0.25$ (9) \nl
HII           & $3.0\pm1.7$ ( 8) & $1.9\pm0.9$ ( 7) & $17.6\pm8.0$ ( 7) & $44.0\pm29.7$ ( 8)  & $0.88\pm0.86$ (8) \nl
Transition    & $3.3\pm2.6$ ( 5) & $3.8\pm0.7$ ( 4) & $ 6.3\pm4.6$ ( 4) & $79.1\pm100.9$ (5)  & $0.44\pm0.45$ (5) \nl
LINER/Seyfert & $3.4\pm2.6$ ( 5) & $7.3\pm4.4$ ( 5) & $ 5.7\pm4.3$ ( 5) & $71.6\pm60.2$ ( 5)  & $0.18\pm0.15$ (4) \nl
\enddata
\tablecomments{
Mean and standard deviation are given for each class. 
Number of objects in each class is in the parentheses.
The standard deviation does not include the errors in each measurement.}
\tablenotetext{a}{
Star formation rate in the central kpc estimated from \Halpha. 
NGC 5194 is excluded from the statistics.}
\end{deluxetable}

\clearpage

\clearpage

\clearpage

\clearpage

\clearpage

\end{document}